\documentclass[useAMS,usenatbib]{mn2e}
\usepackage{epsfig}

\begin{document}

\title{Kelvin-Helmholtz  Instability at the interface of a disc-corona system }

\author[M. Shadmehri and P. Rammos]{Mohsen Shadmehri$^{1,2}$\thanks{E-mail:
mshadmehri@thphys.nuim.ie; }, Perikles Rammos$^{3,4}$\thanks{E-mail: prammos@cp.dias.ie}\\
$^{1}$ Department of Mathematical Physics, National University Ireland, Co Kildare, Maynooth, Ireland\\
$^{2}$ NORDITA, AlbaNova University Center, Roslagstullsbacken 23, SE-10691 Stockholm, Sweden\\
$^{3}$ Dublin Institute for Advanced Studies, 31 Fitzwilliam Place, Dublin 2, Ireland\\
$^{4}$ Section of Astrophysics, Astronomy and Mechanics, Department of Physics, University of Athens, Panepistimiopolis,\\ 157 84 Zografos, Athens, Greece}

\maketitle

\date{Received ______________ / Accepted _________________ }

\begin{abstract}
We study Kelvin-Helmholtz (KH) instability at the interface of a disc and corona system by doing a linear perturbation analysis. The disc is assumed to be thin, however, the corona is considered to be nearly quasispherical because of its high temperature.  Under these circumstances, the interface is subject to the KH instability for a given set of the input parameters. Growth rates of the KH unstable modes are calculated for a wide range of the input parameters. We show that for a certain range of the perturbations, the unstable KH perturbations are growing with time scales comparable to the inverse of the angular velocity of the accretion disc (dynamical time scale). Thus, KH instability at the interface of a disc-corona may have enough time to affect the dynamical structure of its underlying accretion disc by possible exchange of the mass, angular momentum or even energy.  Our linear analysis shows that KH instability may provide a mechanism for such exchanges between a disc and its corona.

\end{abstract}

\begin{keywords}
galaxies: active - black hole: physics - accretion discs
\end{keywords}
\section{Introduction}

The standard theory of the accretion discs \citep*{Shakura1973} has been used widely over recent three decades. Assuming the disc is
radiating like a black body object, we can explain the thermal emission of some of the accretion discs using the
standard theory. However,  non-thermal emission is also detected from some of the accretion discs, namely discs around compact objects or even active galactic nuclei (AGN). One can hardly explain the non-thermal emission based on the standard theory of the accretion discs and for this reason non-standard models are proposed \citep*[e.g.,][]{Kaw2001}. In fact, X-ray of many accreting binaries and AGN imply a hot region around the system and these considerations have motivated some authors to propose a two-component model for these accreting systems, in which a thin  disc is surrounded by a hot and tenuous corona \citep*[e.g.,][]{Haardt91,Nak1993,Kaw2001,Merloni2003}. On the other hand, it is now commonly accepted that magneto-rotational instability (MRI) is the main driving mechanism of the turbulence in the accretion discs \citep*{Balbus1991}. Numerical simulations of MRI show the magnetic field lines may come out of the disc  because of the buoyancy force and generate the magnetic loops outside the disc and dissipate the energy, mainly through the  reconnection mechanism \citep*[e.g.,][]{Miller2000,Liu2002}. Then, the region around the disc is heated significantly because of the magnetic energy dissipation of the loops \citep*[e.g.,][]{Heyv1989,Field1993,Black2000,Kun2004}.  However, there is also another scenario for the formation of a corona around an accretion disc. Some authors believe that the coronae are created by  the radiative pressure driven winds, which produce a hot plasma above an accretion disc \citep*[e.g.,][]{ohsuga2005}.

Since the first phenomenological model of \citet{Haardt91}, the model has been extended in various directions. However, it is still uncertain if the coronae are formed because of the magnetic field or radiation driven winds or even a combination of both mechanisms.  Whatever is the formation mechanism of a corona, the existence of a hot gaseous system above the disc will modify the structure of the underlying accretion disc because of the mass, energy and even angular momentum exchange between the disc and corona. Although the true nature of these exchanges is yet to be understood, one can consider physical arguments to present models for disc-corona systems including possible exchanges. The basic work about the mass exchange between the disc and corona was given by \citet{meyer94}. \citet{czer2000} studied a hot advective corona in pressure equilibrium with the cold disc. In their model, the corona exchanges energy with the disc through the radiative interaction and conduction. In another study, \citet{krolik} presented a model  for the structure of a hot corona by considering the possible effects of thermal conduction. Most of the semi-analytical models for disc-corona systems are following a similar approach \citep[e.g.,][]{czr1999,roz1999,Liu2002,hof2006,Mayer2007}.  Recently, \citet{Mayer2007} presented time-dependent simulations of a two-phase accretion flow around a black hole. They also considered mass and energy exchanges between disc and corona.

In most of the previous studies, for simplicity, it is assumed the corona is rotating with a Keplerian rotational velocity like the disc. Also, the geometrical shape of the corona is not investigated in detail. However, the temperature of the corona is much larger than the disc because the observed high energy cut-off in X-ray spectra usually is about 100 keV. Moreover,  the corona is not thin comparing to its cold disc and the gradient of the pressure in the corona can {\it not} be neglected. Therefore, in our analysis, a corona tends to have  a quasi-spherical shape with a non-Keplerian rotational velocity because of the gradient of pressure. However, in the underlying cold disc the gradient of pressure can be neglected, and it  rotates with Keplerian velocity. In other words, not only there is a significant density contrast between the disc and corona, a velocity drift should exist between the disc and corona.

When two fluids are in relative motion on either side of a common
boundary, the Kelvin-Helmholtz (KH) instability can occur. Thus, we lead to a simple question: Is the interface of a disc and corona system unstable because of KH instability? Our goal is to address this question in this paper. KH instability has already been studied in different astrophysical systems from  the interactions between the solar wind and planetary magnetospheres which are studied based on the KH
instability \citep[e.g.,][]{nagano79} to jets and outflows \citep[e.g.,][]{hardee,rossi} and dusty layer in protoplanetary discs  \citep[e.g.,][]{michikoshi}. Possible instabilities at the interface between an accretion disc and a magnetosphere surrounding the accreting mass have been studied by many authors \citep[e.g.,][]{narayan2004}. It  was shown the magnetospheric interface is generally Rayleigh-Taylor unstable and may also be KH unstable \citep{narayan2004}. But the KH instability at the interface between a disc and its corona has not been studied as far as we know. In the next section, general equations of our model are presented. Analysis of the linear perturbation analysis is done in section 3. A summary of the results  are discussed in section 4.

\section{General formulation}
\subsection{A disc and corona system}
We consider a rotating reference frame at a distance $R$ from the central object (e.g., a black hole or a neutron star) in which the corona is at rest. If we denote the angular velocity of the corona by $\Omega_{c}$, the
continuity and momentum equations in this frame become
\begin{equation}
\frac{\partial\rho}{\partial t}+\nabla . (\rho {\bf v})=0,\label{eq:con}
\end{equation}
\begin{equation}
\frac{\partial {\bf v}}{\partial t}+({\bf v}.\nabla){\bf v}=-\frac{\nabla P}{\rho}-\nabla\Psi - 2 {\bf\Omega}_{c}\times {\bf v} - {\bf\Omega}_{c}\times ({\bf\Omega}_{c}\times {\bf R}),\label{eq:mom}
\end{equation}
where $\rho$, ${\bf v}$, $P$ are  density, velocity and  pressure, respectively. Also, $\psi$ is the gravitational potential due to the central object. The coordinate system $(x,y,z)$ is oriented so that with ${\bf e}_{z}$ parallel to ${\bf\Omega}_{c}$ and ${\bf e}_x$ and ${\bf e}_y$ pointing in radial and azimuthal directions, respectively.

Since the corona is geometrically thick, radial pressure gradients leads to a deviation form pure Keplerian rotation (Mayer \& Pringle 2007).  We can write the $x$ component of the momentum equation (\ref{eq:mom}) as
\begin{equation}
\frac{v_{0y}^2}{R}+2\Omega_{c}v_{0y}+\Omega_{c}^2 R - \frac{1}{\rho_c}\frac{\partial P_c}{\partial R}-\Omega_{K}^2 R=0,
\end{equation}
where $\Omega_K = \sqrt{GM/R^3}$ is the Keplerian angular velocity. Thus,
\begin{equation}
\frac{\Omega_c}{\Omega_K}=\sqrt{1+\frac{\partial P_c / \partial R}{\Omega_{K}^2 R \rho_c}},\label{eq:omeg1}
\end{equation}
or
\begin{equation}
\Omega_c = \frac{\Omega_{K}}{\sqrt{1-(H_c / R)^2\chi_{P}}},\label{eq:omeg}
\end{equation}
where $\chi_P = \partial (\log P_c)/\partial (\log R)$ and $H_c$ is the scaleheight of the corona (Mayer \& Pringle 2007).

\begin{figure}
\epsfig{figure=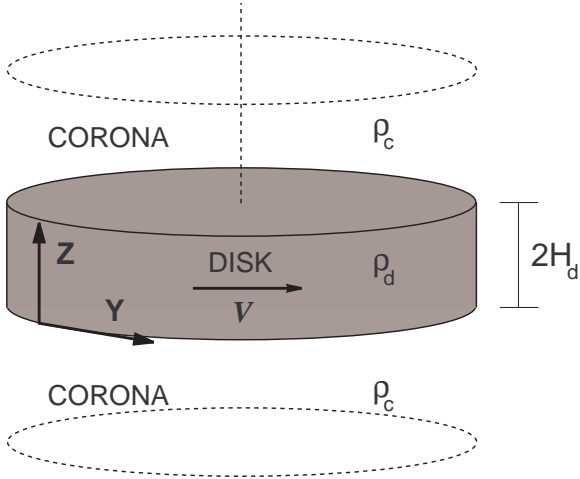,angle=0,width=8cm} \caption{Schematic of the discrete three layer model for a disc and corona system. Generally,  thickness of the corona is much larger than the  thickness of the disc. Although the disc is rotating with Keplerian profile, the rotation velocity of the corona is not Keplerian because of the gradient of the pressure in the corona. In the reference frame rotating with the corona at $\Omega_{c}$, the disc has velocity $V$ at radius $R_0$.}\label{fig:f1}
\end{figure}

Now, we can calculate the rotational velocity of the disc in our rotating reference frame. Since the disc is cold, we can neglect the gradient
of the pressure in the $x$ component of equation of motion (\ref{eq:mom}). Thus,
\begin{equation}
2\Omega_{c} v_{0y} + (\Omega_{c}^2 - \Omega_{K}^2)R=0,
\end{equation}
and using equation (\ref{eq:omeg1}), the above equation can be written as
\begin{equation}
v_{0y}=-\frac{P_c \chi_P}{2\Omega_c R \rho_c}.
\end{equation}
Since $P_c \approx \rho_c c_{s,c}^2$ and $c_{s,c} \approx \Omega_c H_c$, we get
\begin{equation}
v_{0y}=-\frac{1}{2}R\Omega_c (\frac{H_c}{R})^2 \chi_{P}. \label{eq:v0y1}
\end{equation}
Substituting $\Omega_c$ from equation (\ref{eq:omeg}) into equation (\ref{eq:v0y1}), we obtain
\begin{equation}
v_{0y} = \frac{\Theta}{2\sqrt{1+\Theta}} v_{K},\label{eq:v0y}
\end{equation}
where $\Theta = -(H_c/R)^2 \chi_P$ is a nondimensional parameter which depends on the physical structure of the corona. This parameter shows deviation of the rotation of
the corona from  Keplerian angular velocity. Obviously, the parameter $\Theta$ depends on the physical properties of the corona which
are obtained via solving dynamical equations of the corona. Our goal is not to study the detailed structure of corona, though it is a difficult task. Instead,
we consider the parameter $\Theta$ as a free parameter of our model to study possible instabilities at the disc-corona interface. If we neglect the gradient of the pressure of the corona, then $\Theta = 0$ and the disc and corona are both rotating with Keplerian angular velocity. Obviously, in this case, the KH instability is not important. Because both disc and corona are rotating with the same Keplerian angular velocity and there is not drift velocity between them. Thus, in our work, we consider non-zero values for $\Theta$.

\subsection{Linear perturbations}

We assume incompressibility for the analytical calculations and all unperturbed physical quantities are considered constant in
both disc and corona. We perturb the density, the pressure and the velocity of the system as
\begin{displaymath}
\rho (z,y,t) = \rho_1(z) \exp[i(k_{y}y-\omega t)],
\end{displaymath}
\begin{displaymath}
P (z,y,t) = P_1(z) \exp[i(k_{y}y-\omega t)],
\end{displaymath}
\begin{equation}
{\bf v} (z,y,t) = {\bf v}_1(z) \exp[i(k_{y}y-\omega t)].
\end{equation}
After doing mathematical manipulations, the final differential equation for the perturbed $z$ component of the velocity becomes
\begin{displaymath}
{\mathcal D}\{\rho (\omega - k_y v_{0y}) [1-\frac{4\Omega_{c}^2}{(\omega - k_y v_{0y})^2}] {\mathcal D} v_{1z} \}
\end{displaymath}
\begin{equation}
-k_{y}^2 \rho (\omega - k_y v_{0y}) v_{1z} = g k_{y}^2 {\mathcal D}\rho \frac{v_{1z}}{\omega - k_y v_{0y}}, \label{eq:diff}
\end{equation}
where ${\mathcal D} \equiv  d/dz$. In our rotating reference frame, the corona is at rest and the disc has a constant velocity which is given by equation (\ref{eq:v0y}), i.e.
\begin{equation}
V=\frac{\Theta}{2\sqrt{1+\Theta}} v_{K}. \label{eq:V}
\end{equation}

Figure \ref{fig:f1} shows schematic of the discrete three layer model for a disc and corona system.  The differential equation (\ref{eq:diff}) is valid at three layers except at the boundaries $+H_d$ and $-H_d$. Since the density of each layer is assumed to be constant, the general solution of the differential equation (\ref{eq:diff}) becomes
\[v_{1z}(z) = \left\{
\begin{array}{l l}
  C_{1} e^{-q_{1}z}  & \quad \mbox{for $z > H_d$}\\
  C_{2} e^{+q_2 z} + C_{3} e^{-q_{2} z} & \quad \mbox{for $|z|<H_d$}\\
  C_{4} e^{q_{1}z} & \quad \mbox{for $z<-H_d$}\\ \end{array} \right. \]
where $C_1$, $C_2$, $C_3$ and $C_4$ are the constants of integration yet to be obtained from
the boundary conditions appropriate to our system. Note that $q_1$ and $q_2$ are given by the following equations, and are assumed to have a positive real part so as to render the perturbations bounded at infinity. We have
\begin{equation}
q_{1}= \frac{k_y}{\sqrt{1-4\Omega_{c}^2/\omega^2}},
\end{equation}
\begin{equation}
q_{2}= \frac{k_y}{\sqrt{1-4\Omega_{c}^2/(\omega-k_y V)^2}}.
\end{equation}

In general, we may have solutions which are either midplane-symmetric or midplane-antisymmetric. We obtain the dispersion relation for both the cases.

\subsubsection{Even solutions}
If the solutions are even in $z$, we have $v_{1z}(z)=v_{1z}(-z)$ which implies $C_1=C_4$ and $C_2=C_3$. Applying the continuity of the vertical displacements across the interface $z=H_d$ and  integrating  equation (\ref{eq:diff}) over an infinitesimal region around the interface and use standard Gaussian pillbox arguments, we obtain the final dispersion relation as

\begin{displaymath}
\Lambda (\omega - k_y V)^2 [1-\frac{4\Omega_{c}^{2}}{(\omega - k_y V)^2}]^{1/2} \tanh(q_2 H_d)
\end{displaymath}
\begin{equation}\label{eq:even}
+ \omega^2 (1-\frac{4\Omega_{c}^2}{\omega^2})^{1/2}=gk_y (\Lambda -1),
\end{equation}
where $\Lambda=\rho_d / \rho_c$ and $g=g(H_d)=\Omega_{K}^{2}H_{d}$.

\subsubsection{Odd solutions}
In this case, the solutions are midplane-antisymmetric, i.e. $v_{1z}(z)=-v_{1z}(-z)$. So, we have $C_1=-C_4$ and $C_2=-C_3$ and finally

\begin{displaymath}
\Lambda (\omega - k_y V)^2 [1-\frac{4\Omega_{c}^{2}}{(\omega - k_y V)^2}]^{1/2} \coth(q_2 H_d)
\end{displaymath}
\begin{equation}
+ \omega^2 (1-\frac{4\Omega_{c}^2}{\omega^2})^{1/2}=gk_y (\Lambda -1).
\end{equation}

\section{analysis}
\begin{figure}
\epsfig{figure=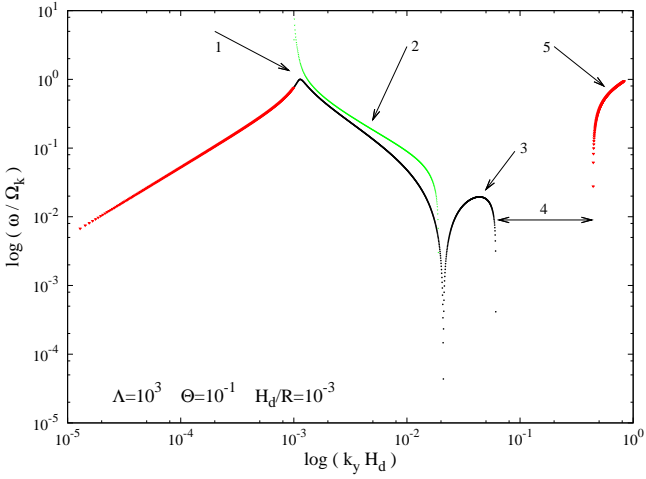,angle=-90,scale=0.32} \caption{Growth rate of the unstable modes versus the wavenumber for $\Lambda =10^3$, $\Theta=0.1$ and $H_d / R = 10^{-3}$. Five different regions  are marked by numbers.  The colours are different roots. }\label{fig:f2}
\end{figure}
\begin{figure}
\epsfig{figure=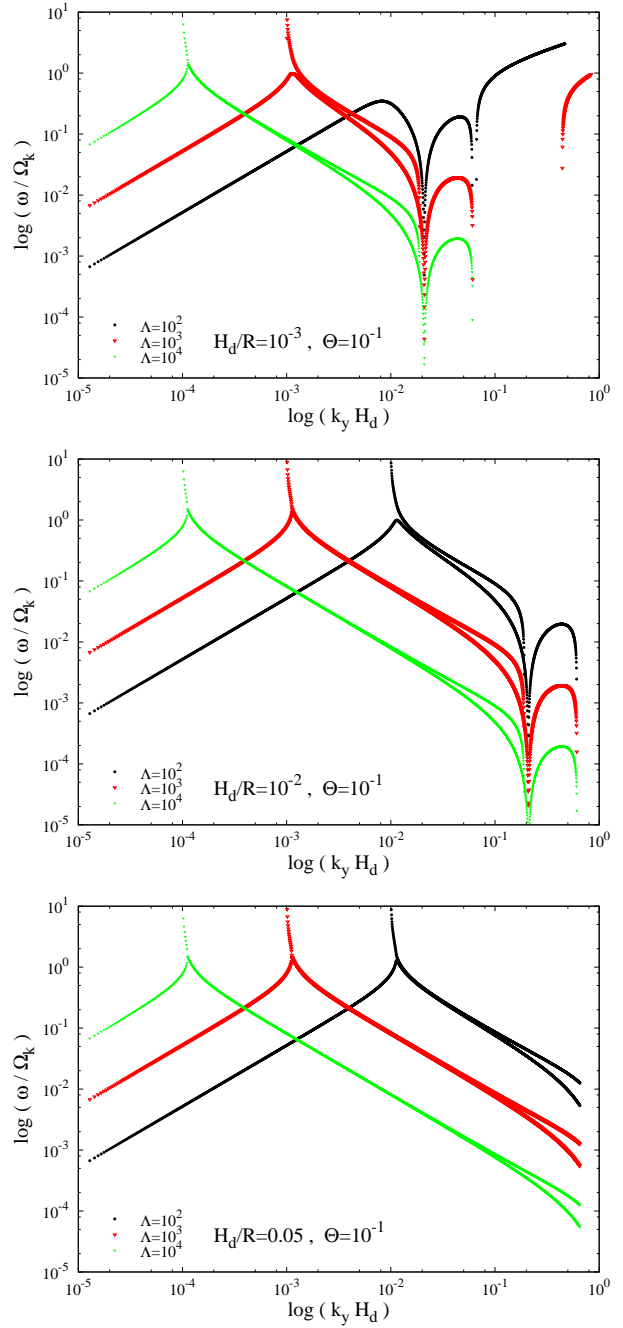,angle=0,width=8cm} \caption{Growth rate of the unstable modes versus the wavenumber for different set of the input parameters which are shown on each plot.}\label{fig:f3}
\end{figure}
\begin{figure}
\epsfig{figure=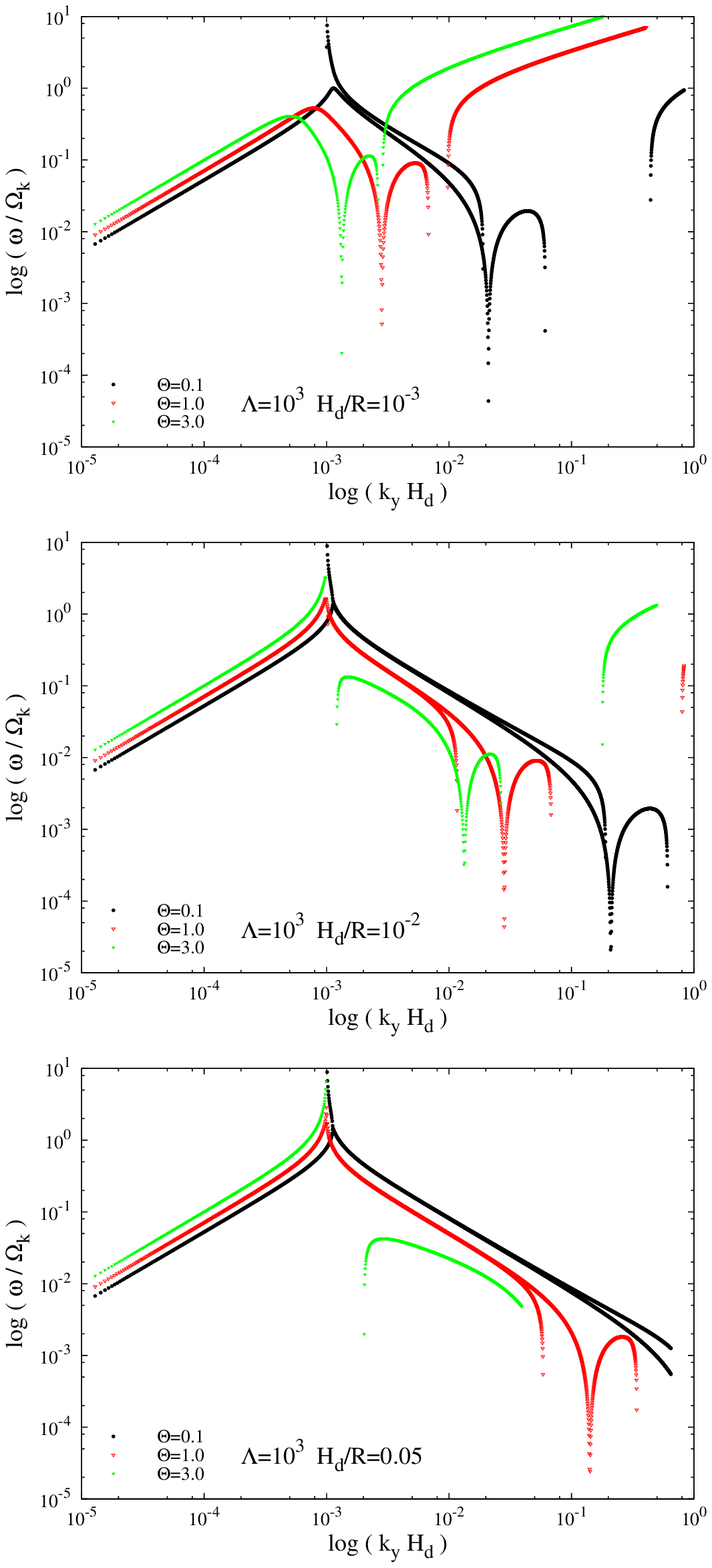,angle=0,width=8cm} \caption{The same as Figure \ref{fig:f3}, but for another set of the input parameters.}\label{fig:f4}
\end{figure}

To study the nature of the solutions of our nonalgebraic equations, one can follow a graphical procedure similar to that in \citet[][p. 498]{chand}. However, we solve the dispersion relation numerically to find unstable modes and their growth rate. Notice that both the even and odd modes have the same instability condition for short wavelengths, i.e. $k_{y} H_{d} \gg 1$.
In this case, the dispersion equation  (\ref{eq:even})
simplifies to the same dispersion relation as the two-layer problem
with rotation studied by \citet{chand}. Thus, our analysis is restricted to
the long-wavelength regime, i.e. $k_{y} H_{d} \ll 1$. In the following we present the dependence of the unstable even solutions
 on the input free parameters of the analysis. However, there a point that we should note. If $\lambda$ represents the wavelength of the perturbations, then we have $\lambda < 2\pi R$. Thus, $k_{\rm y}H_{\rm d} > H_{\rm d}/R$. Although the plots are given for a fixed range of $k_{\rm y} H_{\rm d}$, we should note that the acceptable range is $H_{\rm d}/R < k_{\rm y}H_{\rm d} < 1$ and it depends on the opening angle of the disc.

Now we can introducing the nondimensional growth rate $\Gamma$ and wavenumber $x$ as
\begin{equation}
\Gamma = \frac{\omega}{\Omega_{\rm K}}, x = k_{\rm y} H_{\rm d}.
\end{equation}
Therefore, dispersion equation (\ref{eq:even}) for even solution is written as
\begin{equation}
P_{4} \Gamma^{4} + P_{3} \Gamma^{3} + P_{2} \Gamma^{2} + P_{1} \Gamma + P_{0} =0,
\end{equation}
where the coefficients are
\begin{displaymath}
P_0 = \frac{\left[\Theta ^2 \Lambda  x^3-4 (\Theta +1) (\Lambda -1) x (\frac{H_d}{R})^2\right]{}^2}{16 (\Theta +1)^2 (\frac{H_d}{R})^4}
\end{displaymath}
\begin{equation}
P_1 = -\frac{\Theta  \Lambda  x^3 \left [ \Theta^2 \Lambda x^2 -4 (\Theta +1) (\Lambda -1)  (\frac{H_d}{R})^2\right ]}{2 (\Theta +1)^{3/2}   (\frac{H_d}{R})^3},
\end{equation}
\begin{equation}
P_2 = \frac{3 \Theta ^2 \Lambda ^2 x^4}{2 (\Theta +1) (\frac{H_d}{R})^2}+\frac{4}{1+\Theta}-2 (\Lambda-1) \Lambda  x^2,
\end{equation}
\begin{equation}
P_3 = -\frac{2 \Theta \Lambda ^2 x^3}{\sqrt{\Theta +1} \frac{H_d}{R}},
\end{equation}
\begin{equation}
P_4 = \Lambda ^2 x^2-1.
\end{equation}
There are three input parameters, $\Lambda$, $\Theta$ and $H_{d}/R$. Considering complex dependence of the coefficients on the parameters, it is unlikely to give analytical roots and classify the stability region in the parameter space in closed mathematical forms. However, we solve the above algebraic equation numerically for the parameters  within the acceptable ranges. Roots with non-zero imaginary part are corresponding to the unstable modes. But, we accept only those roots with non-zero imaginary part which are giving $k_y H_d$ less than unity and $q_1$ and $q_2$ with positive real parts.

Figure \ref{fig:f2} shows typical profile of the growth rate versus wavenumber for $\Lambda =10^3$, $\Theta=0.1$ and $H_d / R = 10^{-3}$.  The colours are actually different roots. There are different regions which are labeled by numbers. In region 1, there is only one unstable mode up the wavenumber $k_y H_d = 0.001$. After this wavenumber, we can see two unstable modes (region 2). Growth rate of one mode is slightly larger than the other mode in this region. In fact, the transition between the regions 1 and 2 is determined when the coefficient $P_4$ becomes zero, i.e. $x_{t} = k_y H_d = 1/\Lambda$. As we will show for different values of the input parameters this is the case. In each of the regions 4 and 5, there is only one unstable mode. Also, there is a stability region between regions 4 and 5 for this set of the input parameters. This overall pattern of the growth rate versus the wavenumber is seen for a wide range of the input parameters as we will see. Interestingly, growth rate becomes comparable or even faster than the dynamical time-scale for a certain range of the wavelengths.

In Figure \ref{fig:f3}, we keep the parameter $\Theta$ fixed at $0.1$ but the rest of the parameters are  varied. We can see the transition wavenumber $x_t$ shifts to the smaller values as the ratio $\Lambda$ increase. So, the regions 1 and 2 shift to the left as the ratio of the densities of the disc and corona increases. However, the location of the region 3 is independent of the ratio $\Lambda$, though the growth rate in this region increases with the parameter $\Lambda$. On the other hand, while the stability region 4 vanishes for $\Lambda=100$, the system becomes more stable in this region as the disc becomes more dense in comparison to the corona. When the opening angle of the disc increases, all the regions shift to the right so that the stability region is beyond the acceptable range of the wavelengths of the perturbations.

\begin{figure*}
\epsfig{figure=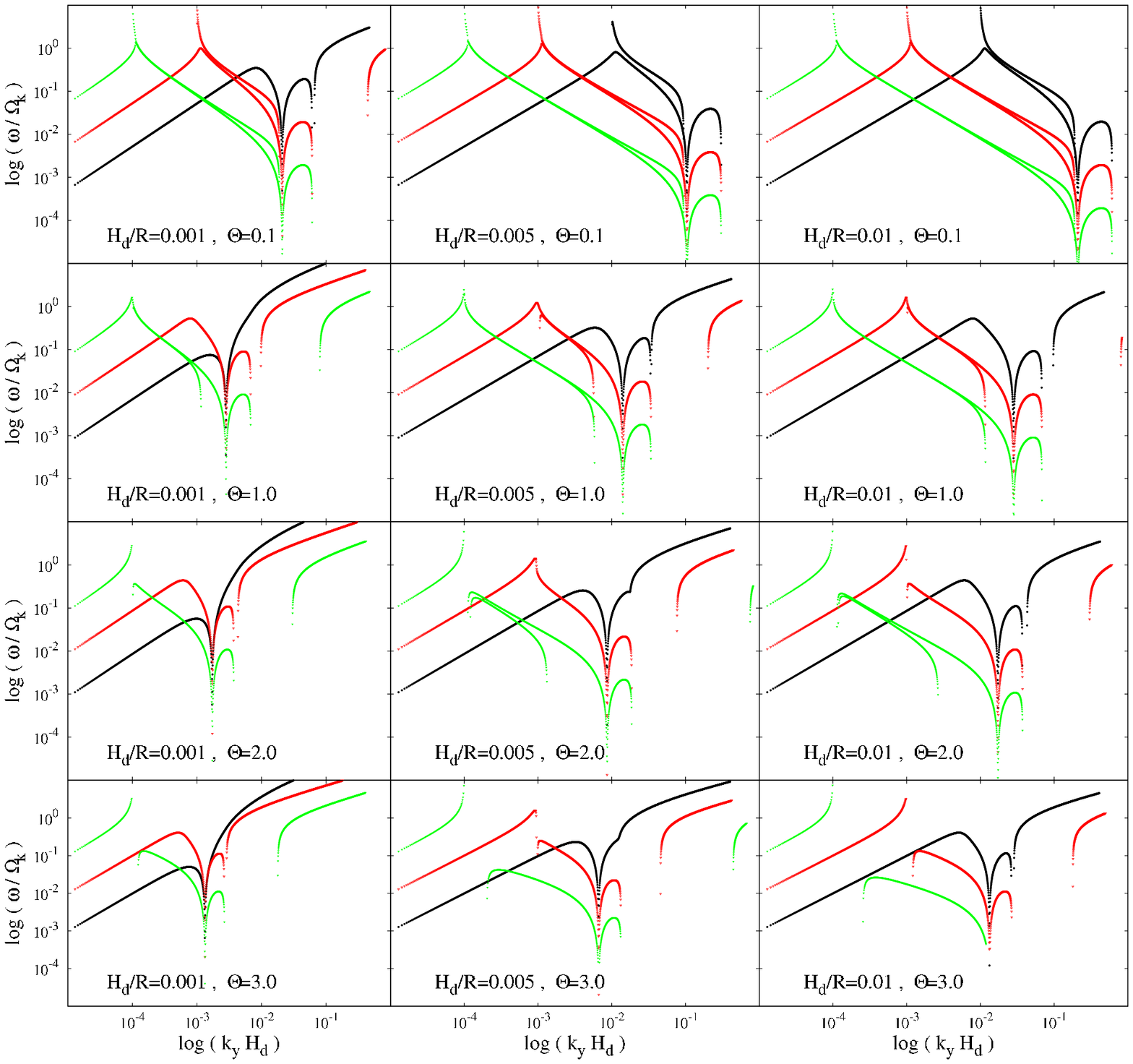,angle=0,scale=0.8} \caption{Growth rate of the unstable modes versus the wavenumber for a wide range of the input parameters. Like Figures \ref{fig:f3} and \ref{fig:f4}, black, red and green lines are corresponding to $\Lambda=10^2$, $10^3$ and $10^4$, respectively. Parameter $\Theta$ varies from $0.1$ to $3$ and the opening angle of the disc, i.e. $H_d / R$ changes from $0.001$ to $0.01$}\label{fig:f5}
\end{figure*}

In Figure \ref{fig:f4}, the ratio $\Lambda$ is kept fixed and the rest of the parameters are changed. As the parameter $\Theta$ increases, not only the curves are shifting to the left, the stability region 4 becomes smaller. Figure \ref{fig:f5} shows growth rate versus wavenumber for a wide range of the input parameters. We can see the overall behavior is similar to the previous figures. In general, increasing $\Lambda$, $H_d /R$ shift the plots to the right and increasing $\Theta$ shift to the left. There are some prominent features, which are altered as the parameters change.  At $x=1/\Lambda$, there is often a very sharp peak comparable to the dynamical time-scale. This special wavelength can be written as $\lambda = 2\pi H_{d} (\rho_{c} / \rho_{d} )$. There is a range of wavelengths for which there are no unstable solutions, and thus the system is stable.

\section{discussion}
We have shown that the interface between a disc and its corona is unstable subject to the KH instability and the growth rate of the unstable modes was calculated. Different time scales are defined for an accretion disc like dynamical and viscous time scales. In most of the analytical models of the accretion discs, dynamical time scale is less than viscous time scale.  For some of the input parameters and a range of the wavelength of the perturbations, the time scale of growing KH unstable modes is comparable to the inverse of the angular velocity of the disc (i.e., dynamical time scale). It means that KH instability has enough time to affect the structure of the disc.   Growth rate of the perturbations versus wavenumbers shows a very complicated profile depending on the input parameters. However, when the wavelength of the perturbations is around $\lambda = 2\pi H_{d} (\rho_{c} / \rho_{d} )$ the time scale of growth is of the order of the dynamical time scale.

It is discussed that the corona may form because of a large scale magnetic field \citep*[e.g.,][]{black2009}. The exact amount of the energy transfer from the disc to the corona by the magnetic flux tubes is yet to be understood theoretically, though there are some estimates based on the physical considerations \citep*[][]{Merloni2003,Kun2004,black2009}. This possible exchange of the energy not only modifies the structure of the corona, but may have significant effect on the structure of the disc as well \citep*{khaj}.  Our analysis proposes another mechanism to be considered for the exchange of the energy or mass between the disc and corona.  Evidently, our analysis does not allow to estimate possible momentum and energy exchanges between the disc and corona because of KH instability. Further numerical simulations are needed to address this issue.

\section*{Acknowledgments}
We are grateful to the anonymous referee whose detailed and careful comments helped to improve the quality of this paper. MS is happy to acknowledge the hospitality of the staff and Axel Brandenburg at NORDITA where parts of this work were done during a research visitor program.

\bibliographystyle{mn2e}
\bibliography{referenceKH}

\end{document}